\documentclass[11pt,twoside]{article}
\usepackage{asp2004}
\usepackage{psfig}
\usepackage{epsf}
\usepackage{graphics}
\usepackage{lscape}
\usepackage{natbib}
\begin{document}
\title{Spectropolarimetry of Core-Collapse Supernovae}
\author{D. C. Leonard}
\affil{Department of Astronomy, California Institute of Technology, Mail Code
  105-24, Pasadena, CA 91125-2400}
\author{A. V. Filippenko}
\affil{Department of Astronomy, University of California, Berkeley, CA
  94720-3411}

\begin{abstract}
We briefly review the young field of spectropolarimetry of core-collapse
supernovae (SNe).  Spectropolarimetry provides the only direct known probe of
early-time supernova (SN) geometry.  The fundamental result is that asphericity
is a ubiquitous feature of young core-collapse SNe.  However, the nature and
degree of the asphericity vary considerably.  The best predictor of
core-collapse SN polarization seems to be the mass of the hydrogen envelope
that is intact at the time of the explosion: those SNe that arise from
progenitors with large, intact envelopes (e.g., Type II-plateau) have very low
polarization, while those that result from progenitors that have lost part
(SN~IIb, SN~IIn) or all (SN~Ib) of their hydrogen (or even helium; SN~Ic)
layers prior to the explosion tend to show substantial polarization.  Thus, the
deeper we probe into core-collapse events, the greater the asphericity seems to
be, suggesting a fundamentally asymmetric explosion with the asymmetry damped
by the addition of envelope material.
\end{abstract}

\section{Introduction}

\begin{figure}[ht!]
\begin{center}

 \scalebox{0.3}{
	\includegraphics{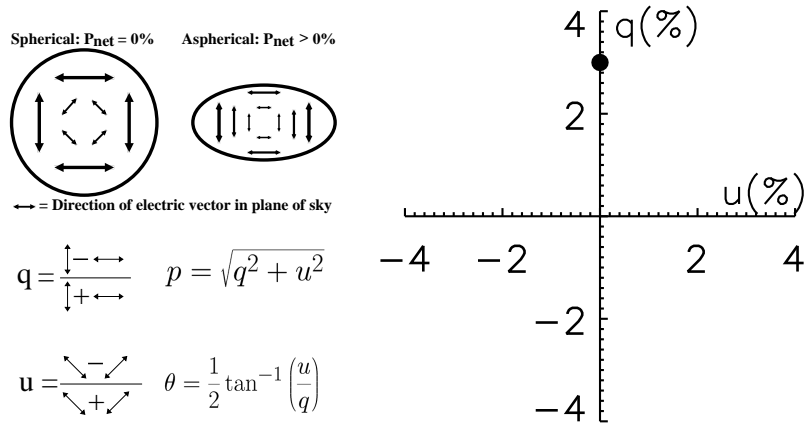}
		}
\end{center}
\vspace*{-0.3in}
\caption{How polarization is produced and measured for an aspherical SN
atmosphere. ({\it Upper left}) Polarization magnitude and direction (in the
plane of the sky) for a resolved electron-scattering atmosphere; for an
unresolved source (i.e., a supernova), only the {\it net} magnitude and
direction can be measured.  Note that the more highly polarized light (longer
arrows) comes from the limb regions in this simple model.  ({\it Bottom left})
By comparing the strengths of the electric vector of the supernova flux at four
different position angles, the normalized Stokes parameters, $q$ and $u$, are
derived, from which the total polarization, $p$, and polarization angle,
$\theta$, may be determined.  As an example, if a young SN actually had the
asymmetry depicted by the ``aspherical'' example in the upper left figure
(approximate axis ratio of 2.0) and was viewed edge-on, a net polarization of
$\sim 3\%$ would result according to the oblate, electron-scattering models of
\citet{Hoflich91}.  This result is indicated by the filled circle in the
$q$-$u$ plane plot shown on the right.
\label{fig:figa} }
\end{figure}

Since extragalactic supernovae (SNe) are spatially unresolvable during the very
early phases of their evolution, explosion geometry has been a difficult
question to approach observationally.  An exciting, emerging field is supernova
(SN) spectropolarimetry, an observational technique that allows the only {\it
direct} probe of early-time SN geometry.  As first pointed out by
\citet{Shapiro82}, polarimetry of a young SN is a powerful tool for probing its
geometry.  The idea is simple: A hot, young SN atmosphere is dominated by
electron scattering, which by its nature is highly polarizing.  Indeed, if we
could resolve such an atmosphere, we would measure changes in both the position
angle and strength of the polarization as a function of position in the
atmosphere.  For a spherical source that is unresolved, however, the
directional components of the electric vectors cancel exactly, yielding zero
net linear polarization.  If the source is aspherical, incomplete cancellation
occurs, and a net polarization results.  In general, linear polarizations of
$\sim 1$\% are expected for moderate ($\sim 20$\%) SN asphericity.  The exact
polarization amount varies with the degree of asphericity, as well as with the
viewing angle and the extension and density profile of the electron-scattering
atmosphere.  Through comparison with theoretical models, the early-time
geometry of the expanding ejecta may be derived (Fig.~\ref{fig:figa}).

Recent interest in SN morphology has been heightened by the strong spatial and
temporal association between some ``hypernovae'' (SNe with early-time spectra
characterized by unusually broad line features) and gamma-ray bursts (GRBs; see
T. Matheson's contribution to these Proceedings).  These associations have
fueled the proposition that some (or, perhaps all) core-collapse SNe explode
due to the action of a ``bipolar'' jet of material
\citep*{Wheeler02,MacFadyen99}, as opposed to the conventional neutrino-driven
mechanism \citep[][and references therein]{Colgate66,Burrows00}.  Under this
paradigm, a GRB is only produced by those few events in which the progenitor
has lost most or all of its outer envelope material (i.e., it is a ``bare
core'' collapsing), and is only observed if the jet is closely aligned with our
line of sight.  Such an explosion mechanism predicts severe distortions from
spherical symmetry in the ejecta.

Largely due to the difficulty of obtaining the requisite signal-to-noise ratio
for all but the brightest objects, the field of SN spectropolarimetry remained
in its infancy until quite recently.  Indeed, prior to our efforts and those of
a few other groups, spectropolarimetry of core-collapse events existed only for
SN 1987A in the LMC \citep[see][and references therein]{Jeffery91} and SN 1993J
in M81 \citep{Tran97}.  The situation has changed dramatically in the last 5
years.  Detailed spectropolarimetric analysis now exists for nearly a dozen
core-collapse events
(\citealt*{Leonard2,Leonard3,Leonard8,Leonard7,Leonard5,Leonard4,Kawabata03,Wang01,Wang03a};
for a review of earlier broadband studies, see \citealt*{Wang96}), and the
basic landscape of the young field is becoming established.  In this review we
discuss the spectropolarimetric characteristics of core-collapse SNe; for a
review of SNe Ia spectropolarimetry, please see the contribution by L. Wang in
these Proceedings.

\vspace*{-0.17in}

\section{Interstellar Polarization}

\begin{figure}[ht!]
\leavevmode
\hspace*{0.2in}
 \scalebox{0.3}{
	\includegraphics{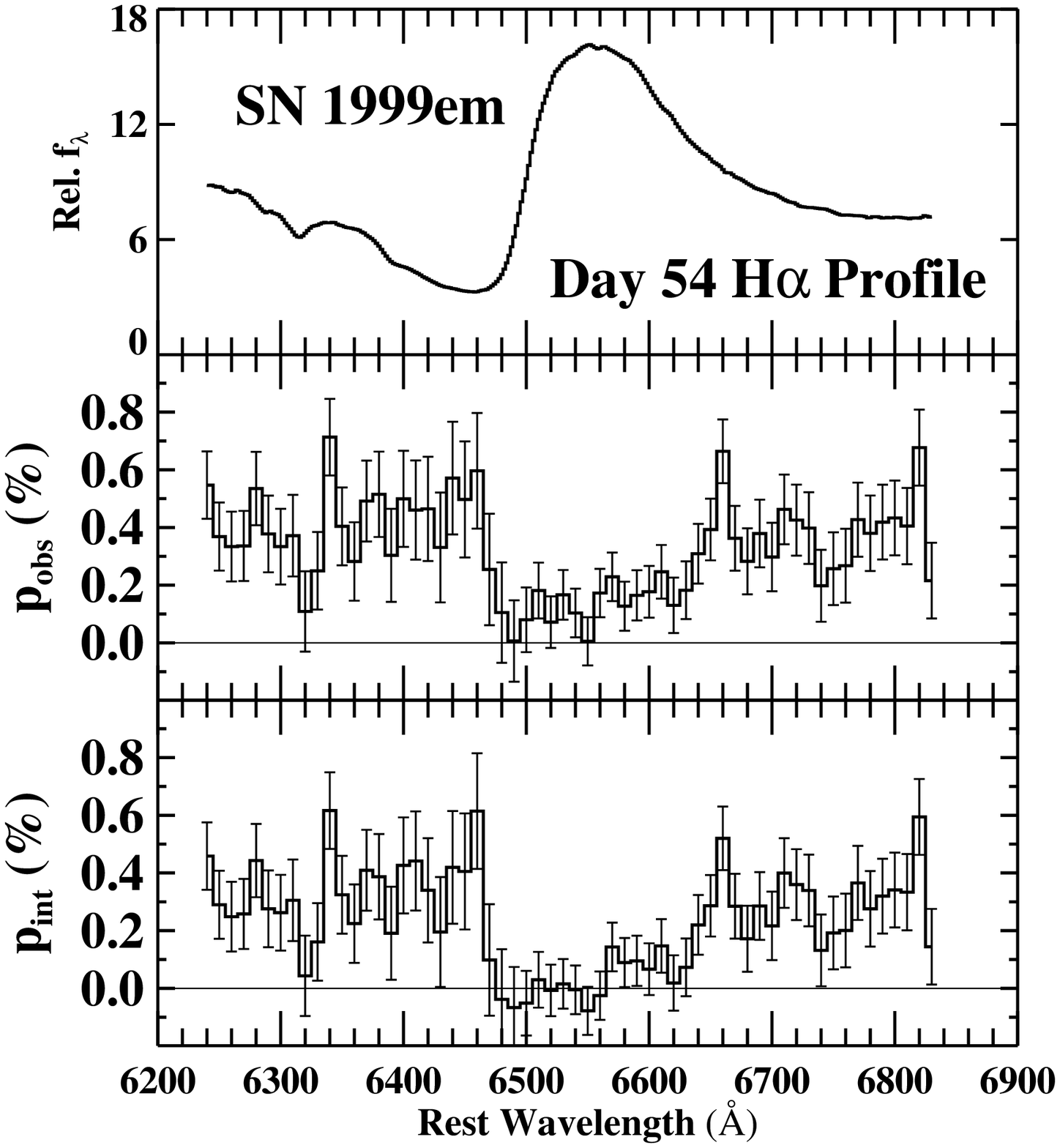}
		}
\hspace*{0.2in}
\rotatebox{90}{
 \scalebox{0.3}{
	\includegraphics{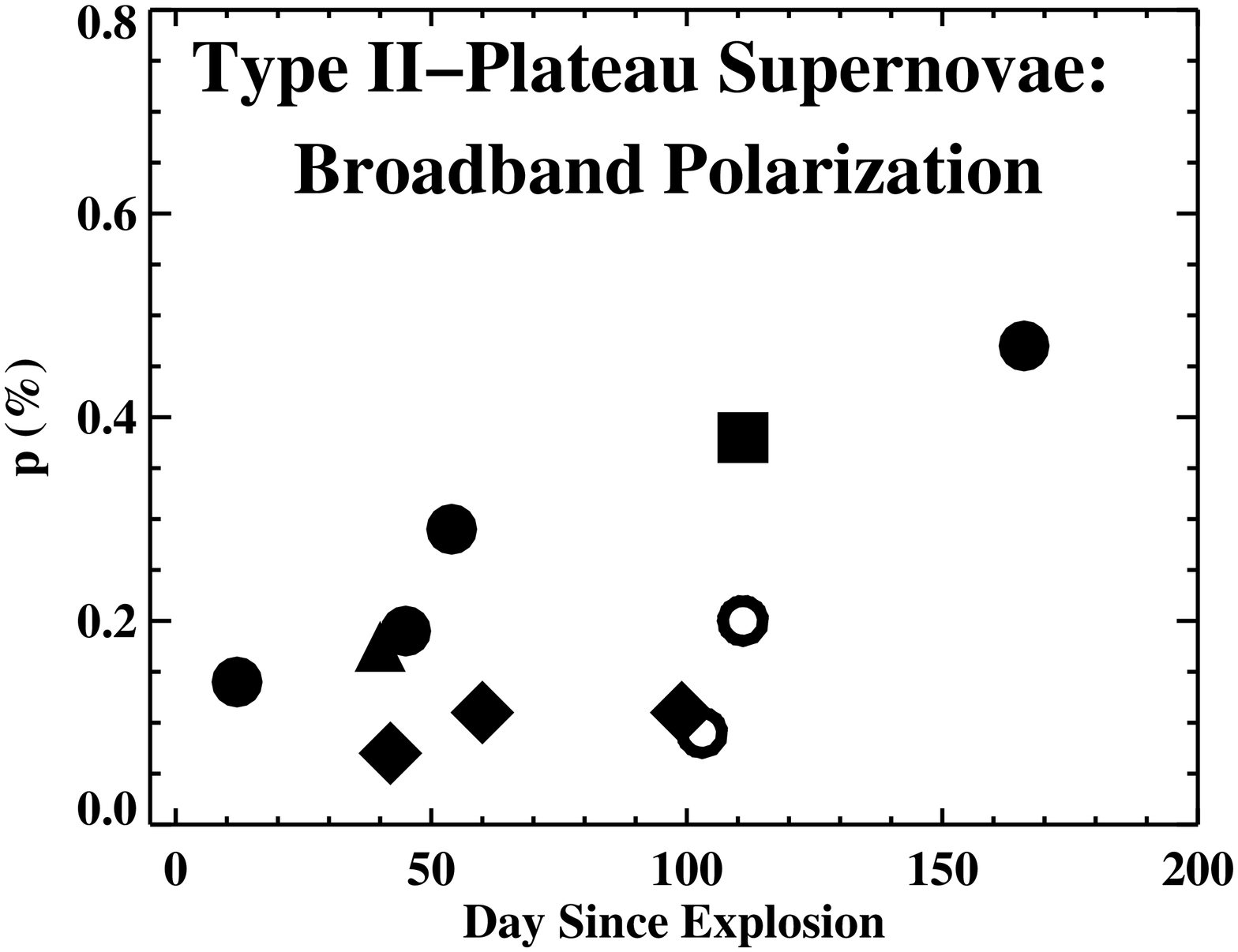}
		}}
\caption{({\it a}, left) Removal of ISP from SN spectropolarimetry.  Total flux
({\it top}, Relative $f_\lambda$), observed polarization ({\it middle}, $p_{\rm
obs}$), and inferred intrinsic polarization ({\it bottom}, $p_{\rm int}$) after
removal of an ISP of $\sim 0.1\%$, for the region around the H$\alpha$ P-Cygni
profile of SN~1999em, 54 days after explosion.  The sharp depolarization at the
location of the strong emission component is a typical signature of SN
spectropolarimetry and is expected from theoretical investigations that
demonstrate that directional information is lost as photons are absorbed and
reemitted in an optically thick line \citep[e.g.,][]{Hoflich96}.  The ISP was
therefore derived by assuming the spectral region near the peak of the
H$\alpha$ line to be intrinsically unpolarized. ({\it b}, right) Broadband
polarization of SNe II-P, after correction for ISP derived under the assumption
of an intrinsically unpolarized spectral region near the H$\alpha$ emission
peak.  The ISP removed, and the plotting symbol used, for each object are as
follows: $0.1\%$ for SN~1999em ({\it closed circle}); 1.0\% for SN~1997ds ({\it
closed triangle}); $5.4\%$ for SN~1999gi ({\it closed square}; for a detailed
study of the extraordinarily high polarization efficiency inferred for the dust
along the line-of-sight in the host galaxy of SN~1999gi, NGC~3184, see
\citealt*{Leonard7}); $0.2\%$ for SN~2001X ({\it closed diamond}); and $1.0\%$
for SN~2003gd ({\it open circle}).
\label{fig:figb} }
\end{figure}

  It is important to first note that a difficult problem in the interpretation
of all SN polarization measurements is proper removal of interstellar
polarization (ISP), which is produced by directional extinction resulting from
aspherical dust grains along the line of sight that are aligned by some
mechanism such that their optic axes have a preferred direction.  The ISP can
contribute a large polarization to the observed signal.  Fortunately, ISP has
been well studied in the Galaxy and shown to be a smoothly varying function of
wavelength and constant with time \citep*[e.g.,][]{Serkowski75}, two properties
that are not characteristics of SN polarization.  This has allowed us to
develop a number of techniques to eliminate ISP from observed SN
spectropolarimetry, the simplest of which is to assume that specific emission
lines or spectral regions are intrinsically unpolarized, and derive the ISP
from the observed polarization at these wavelengths; a demonstration of how
this is accomplished is given in Fig.~\ref{fig:figb}a.

\section{Type II-Plateau Supernovae}

Type II-plateau supernovae (SNe II-P) are the classic variety of core-collapse
events that result from isolated, massive stars with thick hydrogen envelopes
intact at the time of explosion.  The most thoroughly observed SN~II-P is
SN~1999em, for which we obtained rare, multi-epoch spectropolarimetry -- three
epochs during the plateau and one during the early nebular phase.  We found a
very low polarization, $p \approx 0.1\%$, at early times that increased with time
to $p \approx 0.5\%$ by the early nebular phase while maintaining a relatively
fixed polarization angle in the plane of the sky throughout \citep{Leonard3}.
This implies a substantially spherical geometry at early times that may become
more aspherical at late times when the deepest layers of the ejecta are
revealed.  This result provides some support for the jet-induced explosion
models of \citet{Hoflich01}, in which a low but temporally increasing degree of
polarization and, hence, asphericity is predicted for SNe II-P.  A similar
temporal polarization increase was observed for SN~1987A at early times as well
\citep{Jeffery91}.  In fact, for SN~1987A, the expanding ejecta have now
been resolved, and the direction of the observed asymmetry is in accord with the
early-time morphology implied by the photospheric-phase polarimetry data
\citep{Wang02}.

After correcting for ISP, all of the SNe~II-P in our database are found to have
low intrinsic polarization during the plateau (Fig.~\ref{fig:figb}b).  We note
that SN~2001X, for which we obtained multiple spectropolarimetric epochs, does
not show any substantial temporal polarization increase as was seen in SN~1987A
and SN~1999em.  This suggests either a constant degree of asphericity or,
alternatively, a viewing angle (nearly) along an axis of symmetry (e.g., an SN
shaped as a prolate ellipsoid, if viewed pole-on, will show zero polarization
at all times).  The basic result of our investigation into early-time SN~II-P
geometry is that {\it no} event has thus far been found to show large ($\ga
0.5\%$) intrinsic polarization.  However, evidence does exist in some cases for
increasing polarization as one views deeper into the expanding ejecta.

\section{Stripped-Envelope Supernovae}

SNe that have lost a substantial part of their envelopes prior to explosion
display dramatically different polarization signatures than SNe II-P.  These
events tend to be highly polarized, often with significant polarization changes
in magnitude and/or direction in the absorption troughs of the P-Cygni line
profiles.  Well-studied examples include SN~1993J (IIb, $p \approx 1\%$; see
\citealt*{Tran97}), SN~1998S (Type IIn, $p \approx 2\%$; much of the
polarization for this object may result from the interaction of the ejecta with
an asymmetric CSM, see \citealt*{Leonard2,Wang01}), and SN~2002ap (Ic-peculiar,
$p \approx 1.5\%$; see \citealt*{Leonard8,Kawabata03,Wang03a}).  The high
polarization observed for stripped-envelope SNe implies rather extreme
departures from spherical symmetry.

\vspace*{-0.3in}
\begin{figure}[ht!]
\leavevmode
\hspace*{0.2in}
 \scalebox{0.33}{
	\includegraphics{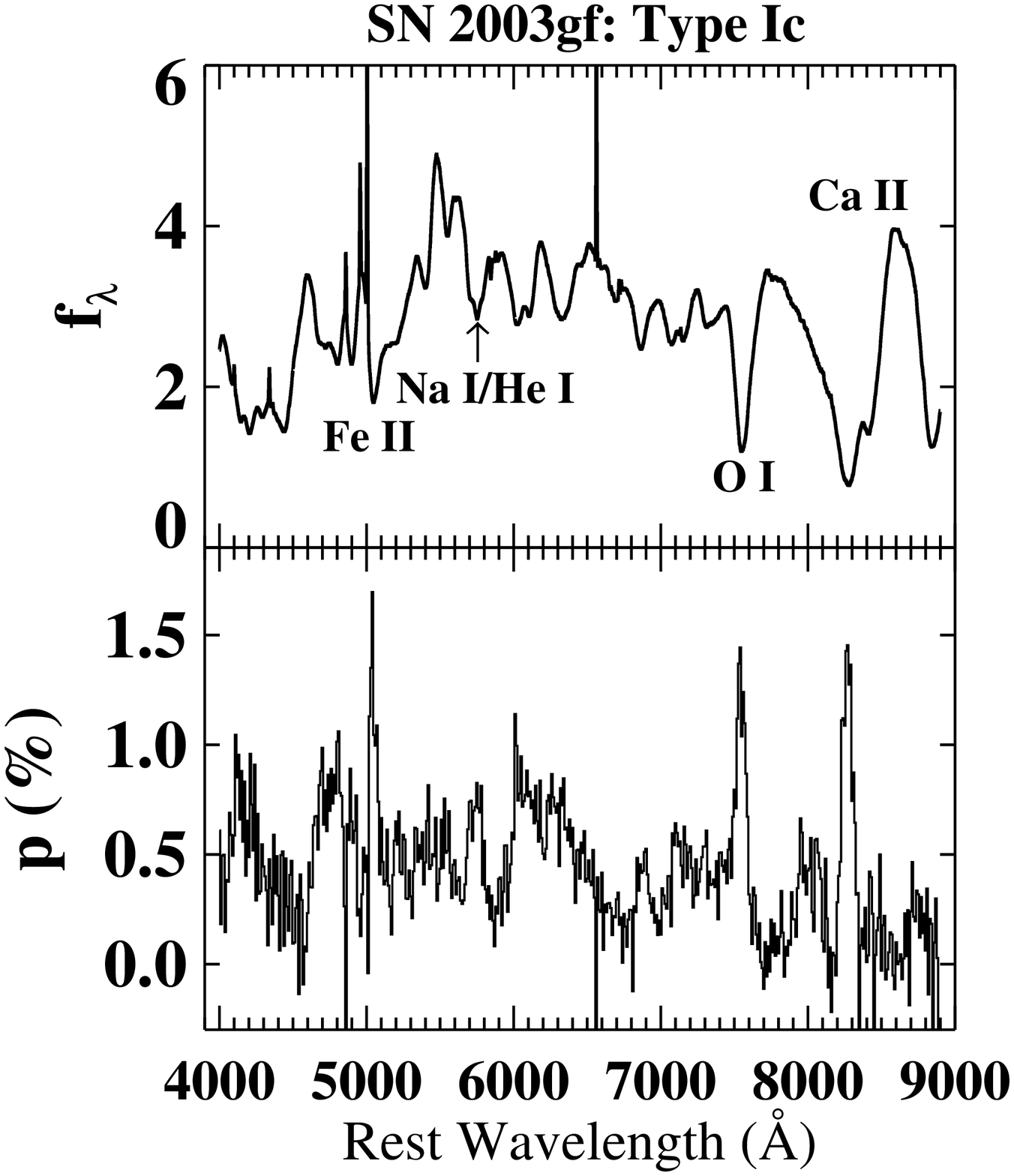}
		}
\hspace*{0.2in}
 \scalebox{0.33}{
	\includegraphics{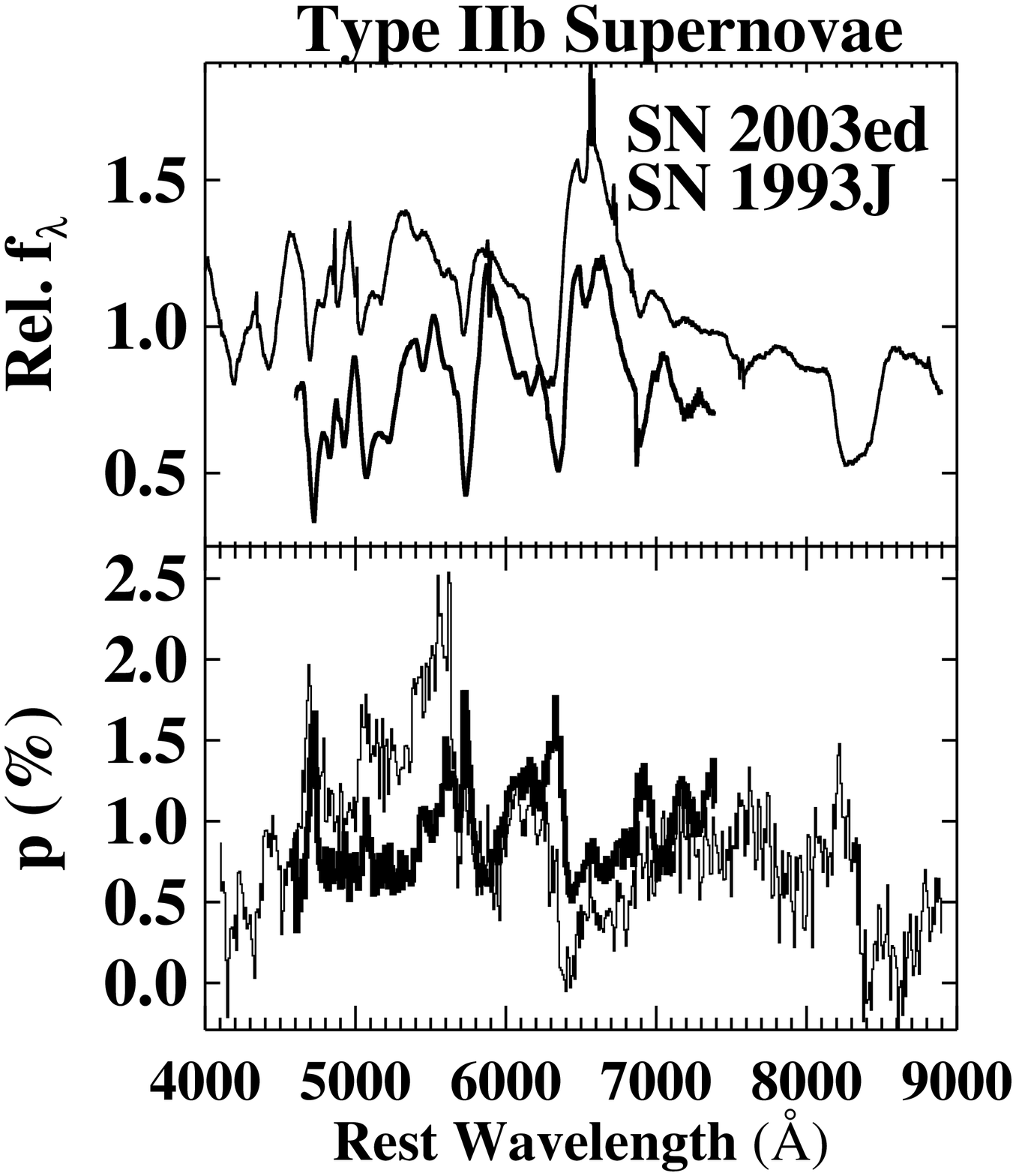}
		}
\caption{({\it a}, left) Spectropolarimetry of SN~2003gf, obtained about 30 days
after explosion.  An ISP of $0.5\%$, derived by assuming zero intrinsic
polarization at the location of the Ca~II near-IR emission peak, has been removed.
({\it b}, right) Spectropolarimetry of SN~2003ed ({\it thin line}, wider spectral
range) and SN~1993J ({\it thick line}, narrower spectral range), taken roughly
$26$ and $34$ days after explosion, respectively.  An ISP of $1.2\%$ has been
removed from the SN~2003ed data (derived by assuming an unpolarized Ca~II near-IR
emission region).  The data for SN~1993J are taken from \citet{Tran97}, and
have 
been corrected for the ISP ($\sim 0.6\%$) derived in that paper.
\label{fig:figf} }
\end{figure}

Two recent examples of stripped-envelope SNe for which we have obtained
spectropolarimetry are SN~2003gf (Type Ic) and SN~2003ed (Type IIb); they are
shown in Figures~\ref{fig:figf}a and \ref{fig:figf}b, respectively.  For
SN~2003gf, note the sharp polarization increases at the location of strong
P-Cygni absorptions. A simple explanation may be that P-Cygni absorption
selectively blocks photons coming from the central, more forward-scattered (and
thus less polarized) regions, thereby enhancing the relative contribution of
the more highly polarized photons from the limb regions (see
Fig.~\ref{fig:figa}).  Clumpy ejecta, or elemental density asymmetries, probably
also play a role in producing these features.

The similarity seen in Figure~\ref{fig:figf}b between the polarization
characteristics of SN~2003ed and SN~1993J at similar epochs is quite
remarkable, especially when coupled with the fact that the only other SN~IIb
for which spectropolarimetry has been obtained, SN 1996cb, also displayed
spectropolarimetric characteristics astonishingly similar to those of SN~1993J
\citep{Wang01}.  The similarity among these three events is somewhat puzzling,
considering the rather wide range of envelope masses that SNe~IIb can be
expected to possess, as well as the effect that random viewing orientations
should have on the resulting spectropolarimetry. As additional data on SNe~IIb
are obtained, it will be interesting to see if the similarities persist.

\section{Conclusion}

The bottom line is that for core-collapse events the closer we probe to the
heart of the explosion, the greater the polarization and, hence, asymmetry.
The small, but temporally increasing polarization of some SNe II-P coupled with
the high polarization of stripped-envelope SNe implicate an explosion mechanism
that is highly asymmetric.  The current speculation is that the presence of a
thick hydrogen envelope dampens the observed asymmetry.  We propose that
explosion asymmetry, or asymmetry in the collapsing Chandrasekhar core, may
play a dominant role in the explanation of pulsar velocities, the mixing of
radioactive material seen far out into the ejecta of young SNe, and even GRBs.

\acknowledgements{D.C.L. is supported by an NSF Astronomy and Astrophysics
Postdoctoral Fellowship under award AST--0401479. A.V.F. is grateful for NSF
grant AST--0307894.}




\end{document}